\documentclass[9pt,twocolumn,twoside]{osajnl}

\journal{ol}
\setboolean{shortarticle}{true}

\title{Exceptional points for photon pairs bound by nonlinear dissipation in cavity arrays}
\usepackage{natbib}
\usepackage{hyperref}

\begin{document}

\newcommand{\e}{{\rm e}}
\newcommand{\rmi}{{\rm i}}
\renewcommand{\Im}{\mathop\mathrm{Im}\nolimits}
\newcommand{\red}[1]{{\color{red}#1}}
\newcommand{\blue}[1]{{\color{blue}#1}}
\newcommand{\green}[1]{\textcolor[rgb]{0,0.7,0.4}{#1}}

\newcommand{\commentSasha}[1]{{\color{red}{\bf Sasha:}\it #1}}
\newcommand{\eps}{\varepsilon}      
\newcommand{\om}{\omega}      
\newcommand{\kap}{\varkappa}      


\author[1,*]{Mark Lyubarov}
\author[2,1,3]{Alexander Poddubny}

\affil[1]{Physics and Engineering Department, ITMO University, St. Petersburg 197101, Russia}
\affil[2]{Nonlinear Physics Centre, Australian National University, Canberra ACT 2601, Australia}
\affil[3]{Ioffe Institute, St. Petersburg 194021, Russia}
\affil[*]{markljubarov@gmail.com}

\begin{abstract}
We study theoretically  the dissipative Bose-Hubbard model describing array of tunneling-coupled cavities with non-conservative photon-photon interaction. Our calculation of the  complex energy spectrum for the  photon pairs reveals 
exceptional points where the  two-photon states bound by nonlinear dissipation are formed.
This improves fundamental understanding of the interplay of  non-Hermiticity and interactions in the quantum structures and can be potentially used for on-demand nonlinear light generation in  photonic lattices.
\end{abstract}

\maketitle

Nonlinear integrated optical circuits are  a promising platform for  quantum information processing
and already reveal fundamental quantum effects~\cite{ICCC_RMP,Peruzzo2010,Solntsev2014,Mittal2017,Barik2018}. One of the simplest interaction effects in a quantum system is a formation of two-boson pairs as a result of repulsive interaction, originating from the Kerr-like nonlinearity~\cite{Mattis1986}, demonstrated for the first time in a  cold atom system~\cite{Winkler2006}. Quantum two-photon effects are also possible in the superconducting resonator networks~\cite{Roushan2016}. Another potential realization is an array of coupled nonlinear cavities, where the  two-photon edge states mediated by interaction have been predicted~\cite{Flach2009,Longhi2013,DiLiberto,Gorlach-2017,DiLiberto-EPJ,Gorlach-H-2017,Salerno}. The system of two interacting photons in a one-dimensional array of nonlinear cavities can be also emulated classically using a two-dimensional array of linear cavities~\cite{Schreiber55,Corrielli2013,Gorlach2018arXiv}. 
While the role of dissipation and dephasing in this system has been studied~\cite{Bello2017},
to the best of our knowledge, the formation of bound two-particle states  has been considered only  due to a  purely Hermitian, conservative, interaction. Spatially modulated losses (or gain) can lead to a dramatic modification of the structure properties and significantly expand the domain of Hermitian systems~\cite{borrmann1950}. In particular, the parity-time  symmetric structures~\cite{Feng2017}, where
losses and gain are symmetrically distributed in space, are beneficial for single mode lasing~\cite{Feng2014,Hodaei2014,Poshakinskiy2016,Pilozzi2017}.  $\mathcal {PT}$-symmetry and non-Hermiticity can be also used to design topological edge states of light, protected from the disorder~\cite{Weimann2016,Leykam2017,Ni2018} and promising for guiding of waves on a chip. 

\begin{figure}[t]
\centering\includegraphics[width=0.3\textwidth]{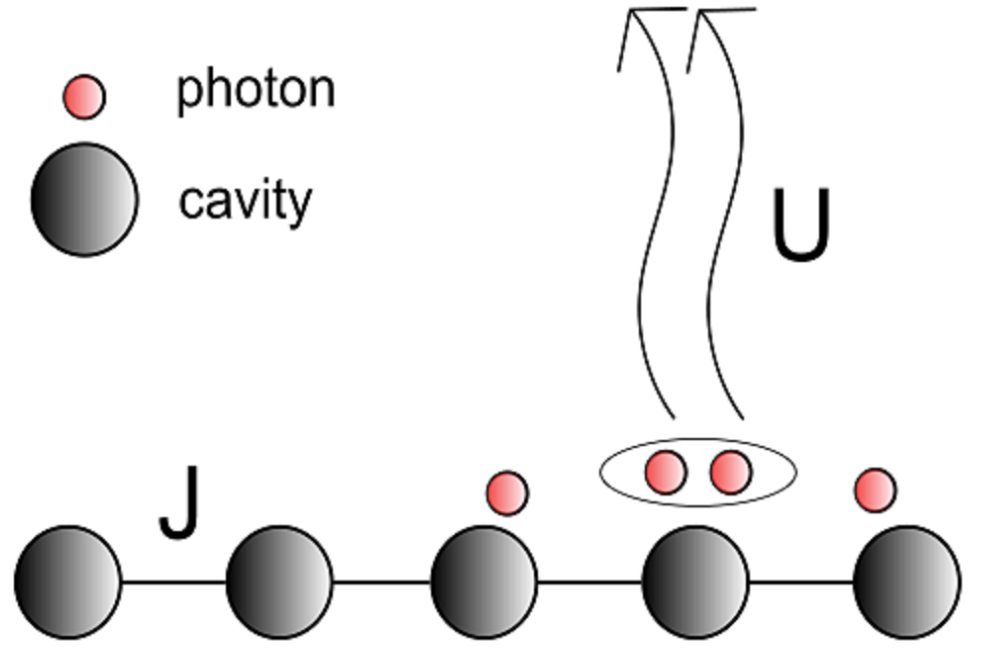}
\caption{Scheme of the array of coupled cavities with the tunneling constant $J$ and dissipative photon-photon interaction $U$.}
\label{fig:1}
\end{figure}

Here, we examine the effect of dissipative interactions on the behavior of photon pairs in the array of nonlinear cavities. Formally, we consider a Bose-Hubbard model with the imaginary interaction term. This can be realized in practice, e.g., by introducing the dissipative Kerr nonlinearity, leading to the absorption of photon pairs located in the same cavity. 
The structure under consideration is schematically illustrated in Fig.~\ref{fig:1}.
It is described by the Hamiltonian 
\begin{multline}
H=H_{0}+U\equiv \hbar\omega_{0}\sum\limits_{j=1}^{N}a_{j}^{\dag}a_{j}^{\vphantom{\dag}}+
J\sum\limits_{j=1}^{N-1}
(a_{j}^{\dag}a_{j+1}^{\vphantom{\dag}}+a_{j+1}^{\dag}a_{j}^{\vphantom{\dag}})\\
-\rmi U\sum\limits_{j=1}^{N}a_{j}^{\dag}a_{j}^{\vphantom{\dag}}(a_{j}^{\dag}a_{j}^{\vphantom{\dag}}-1)\:,\label{eq:Hamiltonian}
\end{multline}
where $a_{j}^{\dag} (a_{j})$ are the photon creation (annihilation) operators, $J$ is the tunneling parameter and $U$ is the dissipative interaction strength.  

We look for the two-photon solutions of the Hamiltonian~\eqref{eq:Hamiltonian} where the wavefunction has the form
\begin{equation}
|\Psi\rangle=\sum\limits_{j,j'=1}^{N}\Psi_{jj'}a_{j}^{\dag}a_{j'}^{\dag}|0\rangle, \label{eq:wavefunction}
\end{equation}
with $\Psi_{jj'}=\Psi_{j'j}$ reflecting the bosonic nature of the excitations. We substitute the wavefunction~\eqref{eq:wavefunction} in the Schr\"odinger equation $E\Psi=H\Psi$ with the Hamiltonian~\eqref{eq:Hamiltonian} and obtain a system of linear equations for the coefficients $\Psi_{jj'}$ in Eq.~\eqref{eq:wavefunction}. 
As such, the interacting two-particle problem in one dimension is exactly mapped to the noninteracting single-particle problem in two dimensions~\cite{Corrielli2013}.
Solution of the system of equations yields the complex energy spectrum of the finite structure. We stress, that the proposed description of the non-Hermitian system by a Schr\"odinger equation with a damping term is approximate. More rigorous model would require the  formalism of master equation for the density matrix or an explicit account for the reservoir describing the losses~\cite{kavbamalas}. However, our simplistic approximation  already allows us to capture the competition between the nonlinear dissipative losses and tunneling-induced  coupling.

\begin{figure}[t]
\centering\includegraphics[width=0.35\textwidth]{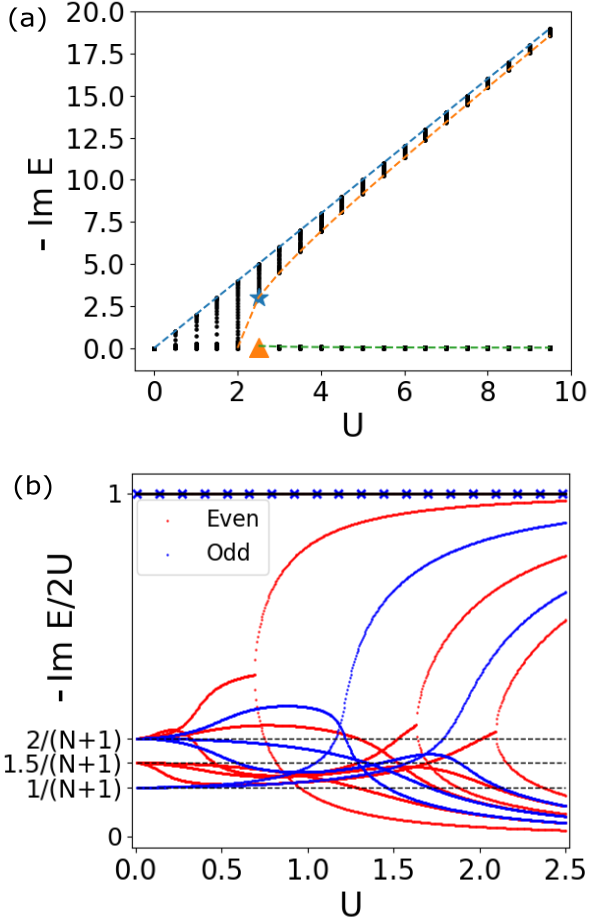}
\caption{Dependence of the imaginary component of the spectrum on the dissipative interaction strength.
(a) Solid circles show the numerical calculation for a finite array; blue, orange and green lines present asymptotic results Eq.~\eqref{eq:as1}, Eq.~\eqref{eq:as2},Eq.~\eqref{eq:as3}, respectively.
Calculation has been performed for $N=30$ cavities and $J=1$.
(b) Same as (a) but in a larger scale  for $N=6$ cavities and normalized to the interaction strength. Red and blue circles correspond to the modes of different parity.
}
\label{fig:2}
\end{figure}

Two types of photon pair eigenstates can be expected. The first type describes the so-called scattering states, i.e. pairs where two photons are  delocalized and quasi-independent from each other. Their  energy is given by a sum of two independent terms corresponding to non-interacting single-photon states. The second type corresponds to bound photon pairs (doublons), where two photons are localized near each other. For the infinite structure the dispersion of these pairs is described by the following equation
\begin{equation}
E(k)=-2\sqrt{4J^{2}\cos^{2}k-U^{2}}\label{eq:dispersion},
\end{equation}
where $k$ is the center-of-mass wave vector, ${\Psi_{jj'}\propto\exp[ik(j+j')/2]}$.
Equation~\eqref{eq:dispersion} was obtained by an analytical continuation of the result in \cite{Winkler2006} to the case of complex photon-photon interaction strength. It indicates that the energy spectrum should exhibit a transition for $U=2J$, and has exceptional points at $k=\pm\arccos U/(2J)$ for $|U|<2|J|$ when the expression under the square root vanishes. We will now verify this by a rigorous calculation for a finite structure. 
Figure~\ref{fig:2}(a) shows the energy spectrum calculated as a function of the interaction strength. It can be seen, that the system exhibits a transition for $U=2J$. For $U<2J$ there is one energy band while for  $U>2J$ the spectrum splits into two separate bands. The upper one corresponds to doublons, bound by a dissipative interaction. Its energies  lie between two asymptotic values
\begin{equation}
E=-2iU, \label{eq:as1}
\end{equation}
and 
\begin{equation}
E=-2\rmi\sqrt{U^{2}-4J^{2}} \label{eq:as2}\:,
\end{equation}
corresponding to the dispersion of the infinite structure~Eq.~\eqref{eq:dispersion} with $k=\pi/2$ (blue curve in Fig.~\ref{fig:2}(a)) and $k=0$ (orange curve), respectively.

The lower band corresponds to quasi-independent photons, i.e. scattering states.  It is instructive to estimate the damping  of these states as a function of  the dissipation $U$ and a number of cavities  $N$. There exist $\sim N^2$ such  states, each state is  distributed over the whole $N\times N $ square, so ${\Psi_{jj'}\propto1/N}$. At the limit of $U\gg 2J$ we have $|E|\ll U$,  and the diagonal cells ($j=j'$) are classically forbidden. 
The imaginary part of energy is responsible for the losses and can be estimated by the perturbation theory. The losses are proportional to  the  part of the wave function located on the diagonal and are due to the mixing of quasi-independent and doublon bands.
The matrix element, responsible for the mixing, is on the order of $J$ and the (imaginary) energy mismatch is $\sim U$. Thus the damping  of quasi-independent photons is $\Im(E)\sim J^{2}/NU$. More precise  asymptotic is provided by an heuristic  expression,
\begin{equation}
\Im E=-\frac{CJ^{2}}{N\sqrt{U^{2}-4J^{2}}}, \label{eq:as3}
\end{equation}
where $C\approx7.66$, that accounts for the transition point at $U=2J$ and well agrees with the numerical calculation in Fig.~\ref{fig:2}(a) (green curve). Interestingly, Eq.~\eqref{eq:as3} indicates that the losses vanish,
$\Im E\to 0$, when we consider an infinite system, $N\rightarrow\infty$. This is because  the  relative contribution of the doublon states, occupying only the 
diagonal of the $N\times N$ Hilbert space, becomes weak.

As expected, the decrease of the interaction strength leads to delocalization and destruction of the doublon modes. However,  the  transition between the regimes with separate doublon and quasi-independent bands ($U\gg 2J$) and mixed bands ($U\ll 2J$) is rather intricate. For small $N$ the transition takes place  at different values of $U$ for different doublon modes. Figure~\ref{fig:2}(b) shows the energy spectrum in the region  $0<U<2.5J$. We consider a small structure with $N=6$ cavities, so that the behavior of all the modes can be traced.
The calculation demonstrates that the fate of the doublons at small $U$ depends on their parity with respect to the transformation 
$j\to N+1-j,j'\to N+1-j'$. The odd modes (blue circles) smoothly change their energy with the interaction strength and delocalize. The even modes (red circles) exhibit exceptional points at different values of $U$. When the  interaction strength becomes weaker than the position of the exceptional point, they coalesce with the scattering states and delocalize. 

More insight into the behavior at $U\ll J$ can be obtained by considering the dissipative interaction as a perturbation.
The eigenstates at $U=0$, describing a pair of non-interacting photons, read
\begin{equation}
|\Psi(k_1,k_2)\rangle=\sum_{j,j'=1}^{N}C_{k_1,k_2}(\sin{k_1j}\sin{k_2j'}+\sin{k_1j'}\sin{k_2j})|j,j'\rangle, \label{eq:wavefunction}
\end{equation}
where $k_i=\pi l_i/(N+1)$, $l_1=1,...,N$, $l_2=1,...,l_1$, and $|j,j'\rangle$ is the state with  photons in cavities $j$ and $j'$. The normalization coefficients are 
 $C_{k_1,k_2}=\sqrt{1+\delta_{k_{1},k_{2}}}/(N+1)$. There exist $N(N+1)/2$ such states, since 
 $ |\Psi(k_1,k_2)\rangle \equiv |\Psi(k_2,k_1)\rangle$. 
 They can be distinguished based on their symmetry and the degeneracy of their spectrum
$
E(k_1, k_2)=2J(\cos{k_1}+\cos{k_2}). 
$
into three groups with $k_{1}=k_{2}$, with $k_{1}+k_{2}=\pi$ and with $k_{1}\ne k_{2}$.

The $N$ states with $k_{1}=k_{2}$ are non-degenerate at $U=0$. The linear-in-$U$ energy correction  is 
\begin{equation}
E=\rmi \langle\Psi(k_1,k_2)|U|\Psi(k_1,k_2)\rangle= - i\frac{3 U}{2(N+1)}\:.\label{eq:E1}
\end{equation}
For the states with $k_{1}\ne k_{2}$ the spectrum is non-degenerate as well and the first order energy correction is equal to 
\begin{equation}
E=-2\rmi U/(N+1)\:.\label{eq:E2}
\end{equation}
There exist $N(N-1)/2$ and $(N-1)^{2}/2-1$ such states for even (odd) $N$, respectively.

The most interesting case is presented by the remaining states with $k_{1}=k_{2}=\pi$. 
They all have energy $E=0$ and the degeneracy $N/2$ ($N/2+1/2$) for even (odd) $N$.
It is convenient to rewrite Eq.~\eqref{eq:wavefunction} for these states as
\begin{equation}
|\Psi(k)\rangle =\frac{\sqrt{2}}{N+1} \sum_{j,j'=1}^{N}\sin{kj}\sin{kj'}[(-1)^{j+1}+(-1)^{j'+1}]|j,j'\rangle\\
\end{equation}
where $k_{1}\equiv k$ and $k_{2}\equiv  \pi - k$. 
In such notation it becomes clear that for even $N$ all these states are odd.
Using this wavefunction we obtain the matrix elements of the perturbation as 
 \begin{equation}
V_{kk'}\equiv \rmi \langle\Psi(k)|U|\Psi(k')\rangle=
\frac{\rmi U}{N+1}(2+\delta_{kk'})\:.
\end{equation}
Solving the secular equation $|V_{kk'}-E\delta_{kk'}|=0$ we finally obtain $N/2$ solutions, including 
\begin{align}
E&=-\rmi U/(N+1)&&\text{$(N/2-1)$ states}\label{eq:E3}\\
E&=-\rmi U&&\text{1 state}\label{eq:E4}\:.
\end{align}
All these modes evolve into doublons with the increase of the interaction strength.
The analytical results Eq.~\eqref{eq:E1},Eq.~\eqref{eq:E2},Eq.~\eqref{eq:E3},Eq.~\eqref{eq:E4}
are shown as horizontal lines in Fig.~\ref{fig:2}(b) and agree with the numerical calculation for $U\ll J$.

\begin{figure}
\centering\includegraphics[width=0.50\textwidth]{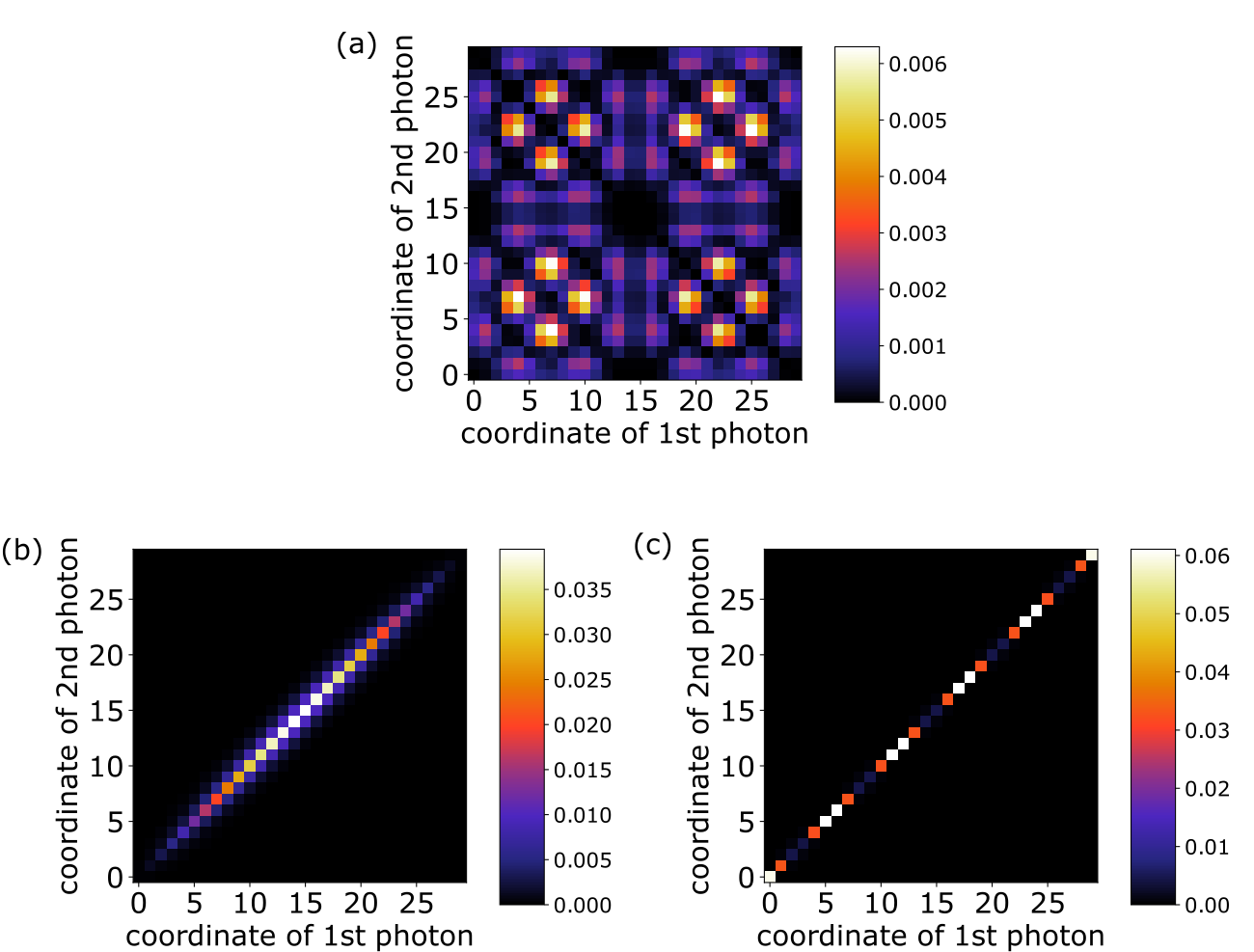}\\

\caption{Illustration of the two-photon wave function for bulk (a) and bound (b), (c) states with $U=2.5J$,
and the energies $E\approx (3.01-0.04\rmi)J$, $E\approx -3.007\rmi J$ and $E\approx -4.89\rmi J$, respectively. 
Other calculation parameters are the same as in Fig.~\ref{fig:2}(a).}
\label{fig:3}
\end{figure}

Figure~\ref{fig:3} shows the wave functions in the real space. Two coordinates in these maps are responsible for the cavity numbers of the two photons, $j$ and $j'$. We consider case with strong interaction, namely $U=2.5J$, so that states from both the doublon band and the band with quasi-independent photons  can be illustrated. In Fig.~\ref{fig:3}(b,c) the two states from the doublon band are presented. Wave function decays away from the diagonal $j=j'$, which means that two photons are localized near each other, as expected from a doublon state. In Fig.~\ref{fig:3}(a) the wave function is distributed over all $N\times N$ square and no localization is observed.

\begin{figure}[t]
\includegraphics[width=0.5\textwidth]{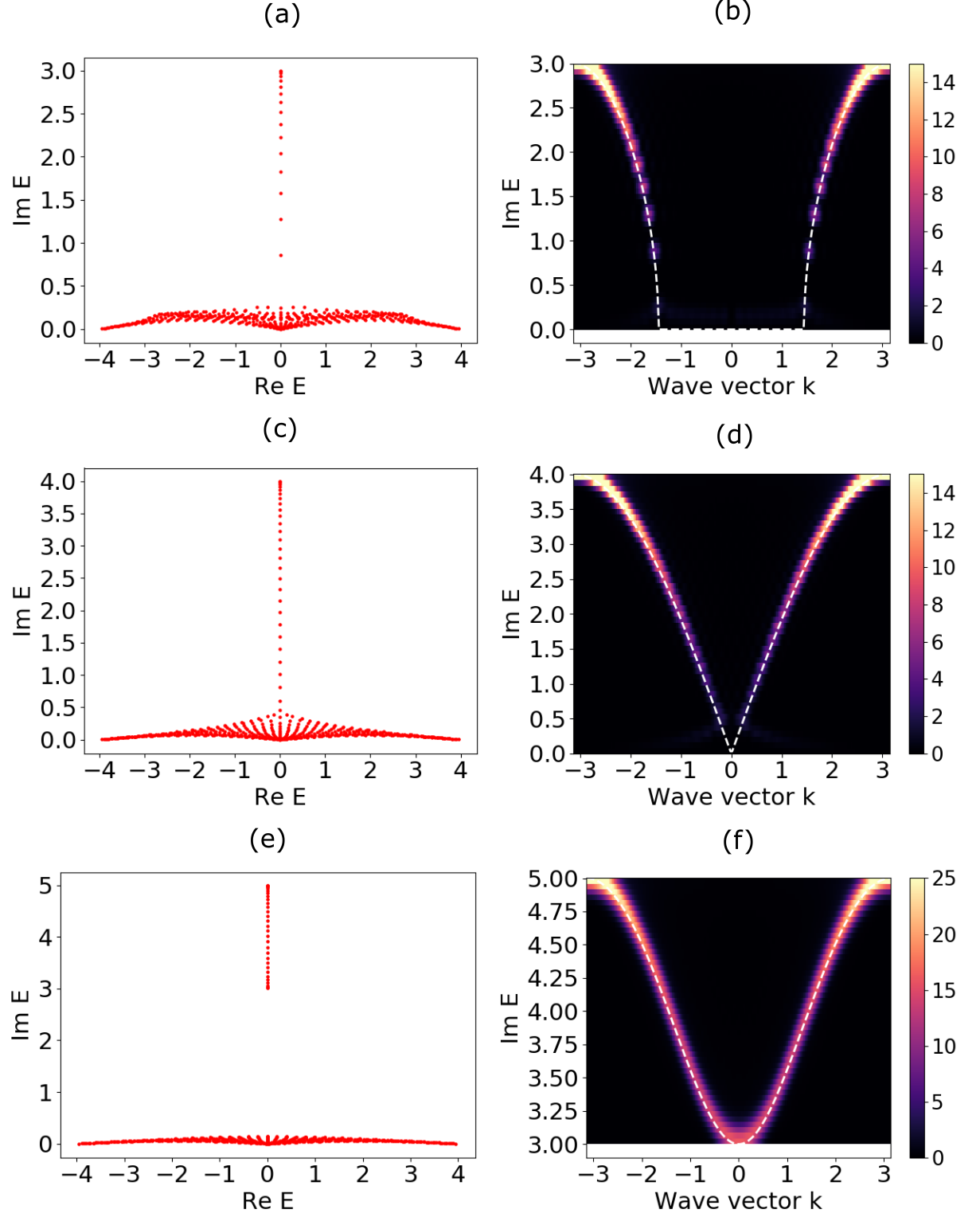}
\caption{(a,c,e)  Complex energy spectrum for $U=1.5J$ (a), $U=2J$ (c), $U=2.5J$ (e).
Panels (b,d,f) show density of states in the reciprocal space for corresponding values of $U$ calculated following Eq.~\eqref{eq:Fourier}. Solid white lines show the analytical dispersion calculated from Eq.~\eqref{eq:dispersion}.
Other calculation parameters are the same as in Fig.~\ref{fig:2}.
}
\label{fig:4}
\end{figure}

As discussed above, the energy dispersion law \eqref{eq:dispersion} of the infinite periodic structure has exceptional points for  $|U|<2|J|$ for certain values of wave vectors. 
For finite $N$ the translational invariance is broken and the  the wave-vector. It is then an instructive question whether these exceptional points in reciprocal space can be directly traced   for a finite cavity array. To this end we perform the Fourier transformation of the  eigenmodes $\Psi^{(\nu)}$ of the finite structure~\cite{Hapala2013}:
\begin{equation}
\widetilde{\Psi}^{(\nu)}_s(k)=\sum\limits_{l=s}^{n}\Psi^{(\nu)}_{l,l-s}e^{-\rmi k(l-\frac{s}{2})}
\label{eq:Fourier}
\end{equation}
to obtain the amplitudes $\Psi^{(\nu)}_{s}(k)$ depending on the photon-photon distance $s$ and the center-of-mass wave vector $k$ and the mode number $\nu$. Next, we introduce  the wave-vector-resolved density of states  $F(E,k)$ for the distribution of photon pairs, located in the same cavity, 

\begin{equation}
F(E,k)=\sum\limits_{\nu}e^{-\frac{|E-E_\nu|^{2}}{\sigma^{2}}}|\widetilde{\Psi}^{(\nu)}_0(k)|^{2}\:.
\label{eq:Fourier_DOS}
\end{equation}
Figure~\ref{fig:4} presents the calculated complex energy spectrum (left column) along with the wave-vector-resolved density of states in the doublon bands (right column) for three values of the dissipative interaction strength, $U=1.5J$, $U=2J$, and $U=2.5J$.  The density of states, extracted from Eq.~\eqref{eq:Fourier_DOS}, agrees well with the  analytical dispersion Eq.~\eqref{eq:dispersion} for the infinite structure, which means that for $N=31$~cavities the doublon wave vector is already a good quantum number. In agreement with the calculation in Fig.~\ref{fig:3} the spectrum demonstrates a qualitative transition at at $U=2J$. While for strong interaction the doublons are defined for all values of wavevector  and their dispersion resembles a typical cosine function [Fig.~\ref{fig:4}(b)], for weak interaction, $U<2J$, the doublons are defined only for larger wavevectors $|k|>\arccos U/(2J)$ and their dispersion features an exceptional point [Fig.~\ref{fig:4}(f)]. The same transition can be traced for the complex energy spectrum: for strong interaction the doublon band is spectrally separated in the complex plane from the band of quasi-independent photons [Fig.~\ref{fig:4}(e)] and for weak interaction the doublons merge with quasi-independent photons with the spectral distance decreasing at large $N$ [Fig.~\ref{fig:4}(a)].

To summarize, we have considered two-photon states in the array of nonlinear cavities with dissipative Kerr-like nonlinearity within the framework of non-Hermitian Bose-Hubbard model. We demonstrate, that the strong dissipative nonlinearity can lead to the formation of well-defined spatially-bound two-photon states. Such doublons have finite lifetime and correspond to photon pairs localized in the same cavity. In case of weak nonlinearity, the formation of the doublon states is also possible for large center-of-mass wave vectors close to the edge of the Brillouin zone. For lower wave-vectors the doublon energy spectrum features interaction-driven exception points, where the doublons merge with the quasi-independent two-photon scattering  states. 

In experiment, if the dissipative interaction is induced by the sum-frequency generation or two-photon absorption processes, the formation of the doublon state would be manifested by short-living traces of  the sum-frequency emission from the same cavity. 
However, for  strong  dissipative interaction the lifetime of the doublons becomes too short,  while for weak interaction they become delocalized. Hence, we expect that there exists an optimal value of the dissipation when the experimental signature of the doublons  would be the strongest. 
We  hope that our results will be useful for emerging studies of quantum, active and topological photonic structures~\cite{Mittal2017,Smirnova2018arXiv,Barik2018}.

We acknowledge useful discussions with M.A.~Gorlach, A. Blanco-Redondo and M.~Di~Liberto.
This work was supported by the Russian Science Foundation (Grant No. 16-19-10538). A.N.P. acknowledges  support from  the Australian Research Council.

\bibliography{TwoPhotNonHerm}

\begin{thebibliography}{10}
\newcommand{\enquote}[1]{``#1''}

\bibitem{ICCC_RMP}
I.~Carusotto and C.~Ciuti, {\protect\JournalTitle{Rev. Mod. Phys.}}
  \textbf{85}, 299 (2013).

\bibitem{Peruzzo2010}
A.~Peruzzo, M.~Lobino, J.~C.~F. Matthews, N.~Matsuda, A.~Politi, K.~Poulios,
  X.-Q. Zhou, Y.~Lahini, N.~Ismail, K.~Worhoff, Y.~Bromberg, Y.~Silberberg,
  M.~G. Thompson, and J.~L. OBrien, {\protect\JournalTitle{Science}}
  \textbf{329}, 1500 (2010).

\bibitem{Solntsev2014}
A.~S. Solntsev, F.~Setzpfandt, A.~S. Clark, C.~W. Wu, M.~J. Collins, C.~Xiong,
  A.~Schreiber, F.~Katzschmann, F.~Eilenberger, R.~Schiek, W.~Sohler,
  A.~Mitchell, C.~Silberhorn, B.~J. Eggleton, T.~Pertsch, A.~A. Sukhorukov,
  D.~N. Neshev, and Y.~S. Kivshar, {\protect\JournalTitle{Phys. Rev. X}}
  \textbf{4}, 031007 (2014).

\bibitem{Mittal2017}
S.~{Mittal} and M.~{Hafezi}, {\protect\JournalTitle{ArXiv e-prints,
  1709.09984}}  (2017).

\bibitem{Barik2018}
S.~Barik, A.~Karasahin, C.~Flower, T.~Cai, H.~Miyake, W.~DeGottardi, M.~Hafezi,
  and E.~Waks, {\protect\JournalTitle{Science}} \textbf{359}, 666 (2018).

\bibitem{Mattis1986}
D.~C. Mattis, {\protect\JournalTitle{Rev. Mod. Phys.}} \textbf{58}, 361 (1986).

\bibitem{Winkler2006}
K.~Winkler, G.~Thalhammer, F.~Lang, R.~Grimm, J.~H. Denschlag, A.~J. Daley,
  A.~Kantian, H.~P. B\"{u}chler, and P.~Zoller, {\protect\JournalTitle{Nature}}
  \textbf{441}, 853 (2006).

\bibitem{Roushan2016}
P.~{Roushan}, C.~{Neill}, A.~{Megrant}, Y.~{Chen}, R.~{Babbush}, R.~{Barends},
  B.~{Campbell}, Z.~{Chen}, B.~{Chiaro}, A.~{Dunsworth}, A.~{Fowler},
  E.~{Jeffrey}, J.~{Kelly}, E.~{Lucero}, J.~{Mutus}, P.~J.~J. {O'Malley},
  M.~{Neeley}, C.~{Quintana}, D.~{Sank}, A.~{Vainsencher}, J.~{Wenner},
  T.~{White}, E.~{Kapit}, H.~{Neven}, and J.~{Martinis},
  {\protect\JournalTitle{Nature Physics}} \textbf{13}, 146 (2016).

\bibitem{Flach2009}
R.~Pinto, M.~Haque, and S.~Flach, {\protect\JournalTitle{Phys. Rev. A}}
  \textbf{79}, 052118 (2009).

\bibitem{Longhi2013}
S.~Longhi and G.~D. Valle, {\protect\JournalTitle{Journal of Physics: Condensed
  Matter}} \textbf{25}, 235601 (2013).

\bibitem{DiLiberto}
M.~{Di Liberto}, A.~Recati, I.~Carusotto, and C.~Menotti,
  {\protect\JournalTitle{Phys. Rev. A}} \textbf{94}, 062704 (2016).

\bibitem{Gorlach-2017}
M.~A. Gorlach and A.~N. Poddubny, {\protect\JournalTitle{Phys. Rev. A}}
  \textbf{95}, 053866 (2017).

\bibitem{DiLiberto-EPJ}
M.~{Di Liberto}, A.~Recati, I.~Carusotto, and C.~Menotti,
  {\protect\JournalTitle{Eur. Phys. J. Special Topics}} \textbf{226}, 2751
  (2017).

\bibitem{Gorlach-H-2017}
M.~A. Gorlach and A.~N. Poddubny, {\protect\JournalTitle{Phys. Rev. A}}
  \textbf{95}, 033831 (2017).

\bibitem{Salerno}
G.~Salerno, M.~Di~Liberto, C.~Menotti, and I.~Carusotto,
  {\protect\JournalTitle{Phys. Rev. A}} \textbf{97}, 013637 (2018).

\bibitem{Schreiber55}
A.~Schreiber, A.~G{\'a}bris, P.~P. Rohde, K.~Laiho, M.~{\v S}tefa{\v n}{\'a}k,
  V.~Poto{\v c}ek, C.~Hamilton, I.~Jex, and C.~Silberhorn,
  {\protect\JournalTitle{Science}} \textbf{336}, 55 (2012).

\bibitem{Corrielli2013}
G.~{Corrielli}, A.~{Crespi}, G.~{Della Valle}, S.~{Longhi}, and R.~{Osellame},
  {\protect\JournalTitle{Nature Communications}} \textbf{4}, 1555 (2013).

\bibitem{Gorlach2018arXiv}
M.~A. {Gorlach}, M.~{Di Liberto}, A.~{Recati}, I.~{Carusotto}, A.~N.
  {Poddubny}, and C.~{Menotti}, {\protect\JournalTitle{ArXiv e-prints,
  1808.05989}}  (2018).

\bibitem{Bello2017}
M.~Bello, G.~Platero, and S.~Kohler, {\protect\JournalTitle{Phys. Rev. B}}
  \textbf{96}, 045408 (2017).

\bibitem{borrmann1950}
G.~Borrmann, {\protect\JournalTitle{Zeitschrift f{\"u}r Physik}} \textbf{127},
  297 (1950).

\bibitem{Feng2017}
L.~Feng, R.~El-Ganainy, and L.~Ge, {\protect\JournalTitle{Nature Photonics}}
  \textbf{11}, 752 (2017).

\bibitem{Feng2014}
L.~Feng, Z.~J. Wong, R.-M. Ma, Y.~Wang, and X.~Zhang,
  {\protect\JournalTitle{Science}} \textbf{346}, 972 (2014).

\bibitem{Hodaei2014}
H.~Hodaei, M.-A. Miri, M.~Heinrich, D.~N. Christodoulides, and M.~Khajavikhan,
  {\protect\JournalTitle{Science}} \textbf{346}, 975 (2014).

\bibitem{Poshakinskiy2016}
A.~V. Poshakinskiy, A.~N. Poddubny, and A.~Fainstein,
  {\protect\JournalTitle{Phys. Rev. Lett.}} \textbf{117}, 224302 (2016).

\bibitem{Pilozzi2017}
L.~Pilozzi and C.~Conti, {\protect\JournalTitle{Optics Letters}} \textbf{42},
  5174 (2017).

\bibitem{Weimann2016}
S.~Weimann, M.~Kremer, Y.~Plotnik, Y.~Lumer, S.~Nolte, K.~G. Makris, M.~Segev,
  M.~C. Rechtsman, and A.~Szameit, {\protect\JournalTitle{Nature Materials}}
  \textbf{16}, 433 (2017).

\bibitem{Leykam2017}
D.~Leykam, K.~Y. Bliokh, C.~Huang, Y.~D. Chong, and F.~Nori,
  {\protect\JournalTitle{Phys. Rev. Lett.}} \textbf{118}, 040401 (2017).

\bibitem{Ni2018}
X.~{Ni}, D.~{Smirnova}, A.~{Poddubny}, D.~{Leykam}, Y.~{Chong}, and A.~B.
  {Khanikaev}, {\protect\JournalTitle{ArXiv e-prints, 1801.04689}}  (2018).

\bibitem{kavbamalas}
A.~Kavokin, J.~Baumberg, G.~Malpuech, and F.~Laussy, \emph{{M}icrocavities}
  (Clarendon Press, Oxford, 2006).

\bibitem{Hapala2013}
P.~Hapala, K.~K\r{u}sov\'a, I.~Pelant, and P.~Jel\'inek,
  {\protect\JournalTitle{Phys. Rev. B}} \textbf{87}, 195420 (2013).

\bibitem{Smirnova2018arXiv}
A.~N. {Poddubny} and D.~A. {Smirnova}, {\protect\JournalTitle{ArXiv e-prints,
  1808.04811}}  (2018).

\end{thebibliography}
\end{document}